\documentstyle[12pt,epsf,rotate]{article}

  \evensidemargin=\oddsidemargin
\def\title#1{\uppercase{#1}}
\def\abstract#1{\centerline{\bf ABSTRACT} \begin{quotation} #1 \end{quotation}}
\def\beginpaper{\relax}
\def\endpaper{\relax}

\def\beginapjbib{\begingroup \section*{References}
         \parskip=.5ex plus 1.0pt 
	 \def\bibitem{\par \noindent \hangindent\parindent
		\hangafter=1}}
\def\endapjbib{\par \endgroup}

\def\be{\begin{equation}}
\def\ee{\end{equation}}
\def\<{\left<}
\def\>{\right>}
\def\sci#1{\ifmmode \times 10^{#1} \else $\times 10^{#1}$ \fi}

\def\alwaysmath#1{{\ifmmode{#1}\else{$#1$}\fi}}

\def\he#1{\hbox{\alwaysmath{{}^{#1}}{\rm He}}}
\def\li#1{\hbox{\alwaysmath{{}^{#1}}{\rm Li}}}
\def\hii{H{\sc ii}~}
\def\dh{\alwaysmath{{\rm D/H}}}
\def\ddash{\,\hbox{to}\,}

\def\Mpc{\;{\rm Mpc}}

\def\etal{{\em et al.\ }}
\def\ltsim{\mathrel{\mathpalette\fun <}}
\def\gtsim{\mathrel{\mathpalette\fun >}}
\def\fun#1#2{\lower3.6pt\vbox{\baselineskip0pt\lineskip.9pt
  \ialign{$\mathsurround=0pt#1\hfil##\hfil$\crcr#2\crcr\sim\crcr}}}

\def\D{{\rm D}}

\begin{document}

\baselineskip=18pt
\rightline{UMN-TH-1432/96}
\rightline{astro-ph/9606156}
\rightline{Submitted to {\em Astrophysical Journal Letters}}

\begin{center}
\title{Implications of a Primordial Origin for the Dispersion \\ in D/H 
in Quasar Absorption Systems}
\bigskip

\vspace{.2in}
Craig J. Copi\footnote{Mailing addresses: \\ \indent Dept. Of Astronomy and
Astrophysics, The University of Chicago, 5640 S. Ellis Ave., Chicago, IL
60637-1433.}

{\it The University of Chicago}\\

Keith A. Olive\footnote{School of Physics and Astronomy, The University of
Minnesota, Minneapolis, MN 55455.}

{\it The University of Minnesota}

and 

David N. Schramm\footnote{NASA/Fermilab Astrophysics Center, Fermi National
Accelerator Laboratory, Batavia, IL 60510-0500.}

{\it The University of Chicago \\ and \\ NASA/Fermilab Astrophysics Center\\
Fermi National Accelerator Laboratory}
\end{center}


\abstract{We consider possible implications if the recent and disconcordant
 measurements of D/H in quasar absorption systems are real and indicate
a dispersion in D/H in these primitive systems. In particular we examine the
option that the D/H abundances in these systems, which are separated on
cosmological scales, are primordial, implying a large scale 
inhomogeneity in the baryon content of the Universe. We show that such large
scale isocurvature perturbations are excluded by current cosmic
microwave background observations. We also discuss the implications of a
smaller (in amplitude) inhomogeneity on the problem of the baryon
density in clusters.}

\newpage

\beginpaper


There has been a recent flurry of activity in measurements of potentially
primordial D/H in quasar absorption systems. If the measured values of D/H are
in fact primordial, then they are of paramount importance to big bang
nucleosynthesis (BBN) since the predicted D/H abundance is a rapidly varying
and monotonic function of the single unknown parameter in the standard model,
namely the baryon-to-photon ratio, $\eta$ (see eg. Walker \etal 1991 and reference therein).  An independent determination of
$\eta$, in addition to enabling a critical test of BBN based on the other light
element isotopes (Reeves \etal 1973), has far reaching consequences on the nature
of dark matter and the composition of galactic halos.  A simple interpretation
based on the recent D/H measurements, however, is complicated by the fact that
they are not uniform.  In fact, they span an order of magnitude in D/H and, as
such taken separately, lend themselves to very different cosmological
interpretations. 
Nevertheless, all measurements are consistent with the basic BBN prediction
that deuterium is primordial and the nucleosynthetic prediction (Reeves \etal
1973; Epstein, Lattimer, \& Schramm 1976) that the primordial D/H ratio is
higher than present day values.  However, the different extra-galactic
measurements, if proven universal, would imply very different chemical
evolution histories with high primordial values indicating large amounts of
stellar processing and lower values indicating perhaps that our Galaxy is
presently receiving primordial infall.
 While it is necessary to be highly cautious at this stage
(as is the case with any new set of data), and indeed any one (or all) of the
reported measurements may not represent the primordial abundance of D/H, here
we entertain the alternate (and even more exotic as we shall see)
possibility that both the high and low D/H measurements are accurate and
uncontaminated. Indeed, it has been suggested (Cardall \& Fuller 1996;
Fuller \& Cardall 1996;
Tytler \& Burles 1996)
that a possible explanation of differing 
measurements of primordial D/H, is the
presence of an inhomogeneity in the baryon number on very
large (cosmological) scales.  We will explore the implications of an
inhomogeneity on the cosmic microwave background, where, if allowed,  we would 
then expect a dispersion
in the height of the Doppler peak to be observable.

In the past two years there have been several measurements of D/H in high
redshift quasar absorption systems.  In the first of these measurements 
(Carswell \etal 1994; Songaila \etal1994) a surprisingly high value of $\log
\dh \simeq -3.92\ddash -3.32$ was reported along the line of sight of
Q0014+813 at $z = 3.32$. The value of
$\eta$ corresponding to this abundance is low, $\eta_{10} = 10^{10} \eta \simeq
1.7$  and was unexpected on the basis of older arguments revolving around the
evolution of  D and \he3 (Walker \etal 1991). Observations of a \he3/H ratio in
planetary nebulae (Rood, Bania
\& Wilson~1992; Rood \etal 1995) indicate the need for at least some
substantial stellar production of \he3 and and thereby necessarily complicate
the evolution of \he3 (see eg.\ Olive \etal 1995; Scully \etal 1996).
 However, \he3 measurements in meteorites
(Geiss 1993), Galactic
\hii\  regions (Balser \etal 1994) and the recent ISM measurement of Gloeckler 
\& Geiss (1996), have cast doubt on traditional \he3 production in low mass
stars (evidenced in the observations of planetary nebulae) and thus the
possibility of high D/H no longer causes conflict on theoretical grounds since
the evolution of \he3 is now suspect. Furthermore, an early report 
of a significantly lower D/H abundance in a different quasar
absorption system (Tytler \& Fan~1995) and the possibility that the
high D/H observation was caused an interloping cloud (Steigman 1994) cast
reserve on the measurements.

 A recent reanalysis (Rugers \& Hogan 1996a) of the absorber at $z=3.32$
revealed in fact two distinct components each with the same high D/H abundance
giving $\log \dh = -3.72 \pm 0.10$ for the average of the two.
The possibility of a hydrogen interloper is thus
 much less likely.  Subsequently,
there have been several other reported measurements of D/H in other quasar
absorption systems. In a system at $z= 3.08$ in the direction of Q0420-388,
Carswell \etal (1996) observe D/H in two close components which both show a high
abundance for D/H (1.0 and 2.5 $\times 10^{-4}$) and when they assume a common
ratio for O/H in the two components (at about 1/10th solar) obtain $\log \dh =
-3.7 \pm 0.1$ consistent with the previous high determination. They
note however, that uncertainties in the H column density could allow values as
low as $-4.70$.  In yet another system at very high redshift, $z= 4.69$,
along the line of sight of BR1202-0725 Wampler \etal (1996) report a D/H
abundance consistent with the other high determinations but quote only an upper
bound of $-3.82$.  Also, Songaila \& Cowie find D/H
$\approx 2 \times 10^{-4}$ in an absorber in Q0956+122, Rugers \& Hogan
(1996b) report $\log \dh = -3.95 \pm 0.54$ in a $z=2.89$ absorber
in GC0636+680. Finally, with respect to high D/H determinations, Rugers
\& Hogan (1996c) have reported a new observation of D/H in a $z=2.80$
absorber again along the line of sight of Q0014+813 with $\log \dh =
-3.73 \pm 0.28$.

There are also quasar absorption systems for which
D/H is measured to be considerably smaller than the high values around $2
\times 10^{-4}$.  In the direction of Q1937-1009, Tytler, Fan \& Burles (1996)
report a value of $\log \dh = -4.64 \pm0.06^{+0.05}_{-0.06}$ in a $z = 3.57$
absorber. Burles \& Tytler (1996) find $\log \dh = -4.60 \pm 0.08 
\pm 0.06$ in a $z=
2.50$ absorber in Q1009+2956. As one can see, there is a clear 
discrepancy between the high and low D/H determinations.  The available
data on D/H in quasar absorption systems is summarized in figure 1.

Implications of either the high D/H measurements or the low D/H measurements
have been discussed relative to BBN and dark matter (Vangioni-Flam \& Cass\'{e}
1995; Hata \etal 1996; Cardall \& Fuller 1996; Fuller \& Cardall 1996;
 Schramm \& Turner 1996). In a
recent analysis of BBN for $N_\nu=3$ based on \he4 and \li7, the two isotopes
which rely least on evolution, Fields
\& Olive (1996) and Fields \etal (1996), determined the range in $\eta$ which
best fits these abundances of these two isotopes as taken directly from
the data. They found $\eta_{10} = 1.8^{+2.0}_{-0.4}$ (for a 95\% CL range).
This corresponds to the range ${\rm D/H} = (0.55 \hbox{--} 2.8)\times 10^{-4}$ 
with a best value sharply peaked at $1.8 \times 10^{-4}$ and is in excellent
agreement with the higher observed D/H seen in {\it some} of the QSO absorbers.
However, such a low value of $\eta$ becomes more
problematic for the evolution of \he3 which appears to be relatively flat over
the last 5 Gyr history of the galaxy (Turner \etal 1996).  We emphasize that
when the production of \he3 in low mass stars is included in chemical evolution
models (Olive \etal 1995; Galli \etal 1995; Scully \etal 1996; Dearborn,
Steigman \& Tosi 1996) even higher values of $\eta$ (lower values of
primordial D/H) are problematic and the issue of the evolution of \he3 becomes
a question for stellar evolution (Charbonel 1994, 1995;  Hogan 1995; Wasserburg,
 Boothroyd, \& Sackman 1995; Weiss \etal 1995; Boothroyd \& Sackman 1995;
Boothroyd \& Malaney 1995).

The low D/H determinations also can be problematic for BBN using a straight
interpretation of the data.  Although the low D/H does relax to some extent
the problem of \he3 evolution (Scully \etal 1996), there may be problems with
the abundances of the other light elements. Because the observationally
determined values with their published systematic errors of \he4 and \li7 fit
very well with the high D/H measurements, the combined likelihood of all
three isotopes when convolved with the BBN calculations is extremely high
(Fields \etal 1996).  To get a comparable likelihood,
the \he4 determinations would have to be systematically low by 0.015, that is
as opposed to the value $Y_p = 0.234 \pm 0.003$ for the \he4 mass fraction
(Olive \& Steigman 1994; Olive \& Scully 1996), the central value for $Y_p$
would need to be raised to 0.249, a value which is high even when a generous
assessment of the systematic errors is made (Copi, Schramm, \& Turner 1995).
 Furthermore, \li7/H would
need to be as high as 5 $\times 10^{-10}$ requiring as much as a factor of 3
depletion which may be problematic (Steigman \etal 1993).

There is of course the possibility that either (or both) the high or low D/H
determinations are not an accurate indication of primordial D/H as it pertains
to BBN.  Potential problems with the high D/H observations have been outlined
by Burles \& Tytler (1996) and include uncertainties in the hydrogen column
density which can lead to a serious over estimate of the D/H ratio. This is  a
chief uncertainty in the case of Q0420-388 and one of the reasons that this
observation is consistent with the low D/H measurements.  The absorber in
BR1202-0725 has relatively high metallicity ([O/H] = 0.3) which casts doubt on
its primordial nature (though this would indicate a potentially higher D/H) and
the lack of metal lines in the spectra of Q0014+813 casts some
uncertainty on the accuracy of these abundances.

The low D/H measurements are also subject to considerable uncertainty.
As pointed out by Rugers \& Hogan (1996c), the gas mass covering the quasar
image is very small (less than a solar mass). Thus the ejecta of a single
intermediate mass star (3--8 M$_\odot$) which is both deuterium free and
heavy element free can greatly affect the measurement of D/H.  Clearly such an
effect would lower the D/H ratio in an absorber.  In addition, the abundances
in the two recent low D/H systems of Tytler, Burles \& Fan (1996) and Burles \&
Tytler (1996) are based on blended lines where it was assumed that the D/H was
uniform in the two components. Other assumptions could increase the implied
D/H ratio. It fact, it has been recently argued (Wampler 1996), 
that alternative cloud models could raise the low D/H abundances by a 
factor of 3--6.  A factor of 6 would make them compatible (within errors) 
of the high D/H measurements. 
It is clear that the most prudent course of action is to treat with
some reserve
 all of the QSO absorption system measurements of D/H\@. Hopefully, with
increased statistics, a better understanding of the measurements and the D/H
will be available.  One would further expect that either the observers finding
high D/H should begin to see some systems with low D/H, vice versa, or both. 

Given the appropriate amount of caution regarding these measurements, let us 
now assume that {\em all} of the D/H measurements in quasar absorption
systems are in fact, accurate.  Therefore we may ask, can the value of
primordial D/H differ in these systems.  In the remainder of this letter, we
would like to detail the consequences of this assumption.

As we stated earlier, the BBN prediction for primordial D/H is a very rapidly
changing monotonic function of the baryon-to-photon ratio, $\eta$.  As such,
any measurement with some degree of confidence of primordial D/H, can in
principle very accurately pin down the value of $\eta$.  If there is a real
dispersion in primordial D/H as would be the case if both the high and low D/H
measurements were accurate, then we have evidence for the first time of a real
inhomogeneity in the baryon number on large scales. In addition, the amplitude
of the fluctuations producing these inhomogeneities must be large (${\cal O}
(1)$) at the time of BBN as the D/H dispersion corresponds to values of $\eta$
from $\sim 1.5$ to $\sim 7.0$. The various QSO absorption systems in which the
D/H measurements are made are all at large redshift (ranging from $z =
2.5\hbox{--} 4.7$) and are thus separated by cosmological distance scales.  We
stress that 
inhomogeneities on scales as large as this are very different from the
adiabatic baryon inhomogeneities inspired from the QCD phase transition (see
eg. Malaney \& Mathews 1993; Thomas \etal 1994) and from the small scale
baryon isocurvature fluctuations (Kurki-Suonio, Jedamzik, \& Mathews~1996)
that give rise to inhomogeneous BBN\@.  In these cases the inhomogeneities were
produced on small scales (sub-horizon at the time of BBN), while the
inhomogeneities we are considering must be nearly horizon scale today, meaning
that these are inhomogeneities on scales much larger than the horizon at the
time of BBN.  That is, within a horizon at the time of BBN, we are still
dealing with homogeneous BBN. Nucleosynthesis with large scale inhomogeneities
has been considered recently (Copi, Olive, \& Schramm 1995; Jedamzik \& Fuller
1995).  These studies focused on inhomogeneities that have mixed and would not
lead to inhomogeneities in the observed abundances.

We would also like to emphasize the implications of a large scale
baryon inhomogeneity of the type we have been discussing on standard
BBN.  The abundance
of a single light element isotope which can be associated with its primordial
value is enough to constrain BBN and determine a value for $\eta$.  
To test the theory, we need the abundances of two or more light element
isotopes.  For example, this can be and is achieved using \he4 and \li7 (Fields
\& Olive 1996). If the Universe were homogeneous in baryon number, then a
primordial measurement of D/H in quasar absorption systems would lead to strong
test of the theory.  However, if the dispersion in D/H in these systems is real
and indicates a real large scale baryon inhomogeneity, then unless a second
isotope (eg. \he4 or \li7) can be observed in the {\em same} absorption system,
the D/H measurements will yield information regarding the baryon-to-photon ratio
in those systems, in addition to information on the inhomogeneity.
These measurements alone could not be used to test BBN as they could not be
directly compared to the predictions based on \li7, which is observed in our
own Galaxy, or \he4, which though is observed in external galaxies, these are
all relatively local (ie.\ they are all at very low redshift).

\bigskip 

One important and viable mechanism for the production of the baryon asymmetry
of the Universe is realized in supersymmetric extensions of the standard model
of electroweak interactions (Affleck \& Dine 1985; Linde 1985a; Ellis \etal
1987).  The minimal supersymmetric standard model contains many additional
particle degrees of freedom over the standard model, corresponding to the
supersymmetric partners of ordinary particles.  The potential
for the scalar fields contains directions for which the potential is
perfectly flat, that is, certain combinations of these fields are allowed to
take arbitrarily large vacuum expectation values at little (or none, if
supersymmetry is unbroken) cost in energy. During inflation, De Sitter
fluctuations drive these scalar fields to large expectation values and the
subsequent evolution of these fields (which store baryon and lepton number)
produces a baryon asymmetry.  The final baryon asymmetry is in general quite
model dependent as it depends on quantities such as the expectation value
produced by inflation, the inflaton mass, the grand unified mass scale, and
the supersymmetry breaking mass scale.

In the course of the evolution of the scalar fields, sfermion density
fluctuations are produced (Ellis \etal 1987; Yokoyama 1994) which lead to
isocurvature fluctuations in the baryon density 
with an amplitude which is given by (Linde 1985b)
\begin{equation}
{\delta n_B \over n_B} = {\delta \rho_\phi \over \rho_\phi} \simeq
{\delta \phi(k) \over \phi} \simeq
(H / \phi_o)(k/H)^{{\tilde m}^2/3H^2}
\end{equation}
  where $H$ is the Hubble parameter during
inflation, $\tilde m$ is the sfermion mass, and $\phi_o$ is the vacuum
expectation value of the sfermion fields, $\phi$, produced during inflation.
Thus, isocurvature baryon number fluctuations are produced with an amplitude
which depends primarily on  the ratio $H/\phi_o$ and may take values from
$\sim 10^{-8}$ to $\sim 1$. In order to have an impact on the D/H abundances,
$\delta n_B/ n_B$ must take values of order 1, however, smaller values though
not important for D/H, may still
have significant cosmological consequences for the density of baryons in 
clusters of galaxies. We will return to this point below, but 
for now we will assume 
$\delta n_B/ n_B \sim 1$.

The overall baryon to entropy ratio is given by (Ellis \etal 1987)
\begin{equation}
{ n_B \over s} = {\phi_o^4 m_\psi^{3/2} \over M_G^2 M_P^{5/2} {\tilde m}}
\sim 10^{-11}
\end{equation}
for $\phi_o \sim 10^{-6}M_P$, an inflaton mass, $m_\psi \sim \phi_o$, a GUT mass
of order  $10^{-3} M_P$, and a sfermion mass, ${\tilde m}$,
 of order 100 GeV. First it is
important to stress that in models such as these,the baryon density
 $\rho_B \ll \rho_{\rm total}$
so that even though
$\delta
\rho_B/\rho_B$ may be of order unity (if $H \sim \phi_o$), $\delta
\rho_B/\rho_{\rm total} \sim 10^{-9}$ is very small at the time the sfermion
oscillations have decayed, and
$\rho_{\rm total}$ is the total energy density which is dominated by the
dynamics of inflation. 
Second, the spectrum of isocurvature perturbations in the baryons is very nearly
flat, since ${\tilde m}/H \ll 1$, is present on exponentially large
scales, and is completely independent of the adiabatic 
fluctuations produced by inflation.  Note that adiabatic perturbations 
do not influence BBN since
$\delta\eta/\eta = 0$. (More accurately, $\delta(n_B/s)=0$ for adiabatic
perturbations.) Thus, the overall large scale structure believed to have
been seeded by inflation remains a flat adiabatic spectrum of density
fluctuations.  Baryons, however have in addition an isocurvature component with
an amplitude which may be of order unity.
Such a model provides a plausible physical origin for baryon inhomogeneities
on very large scales. 

To explain the dispersion in D/H as seen in different quasar absorption
systems we require large scale inhomogeneities at the time of nucleosynthesis.
Since these quasar absorption systems are observed along different lines of
sight the isocurvature power spectrum must have significant power on angular
scales $\theta \gtsim {\cal O} ({\rm few})\;{\rm degrees}$. Furthermore they
must be created with a large amplitude since they must be in place at the
onset of BBN to generate the observed dispersion in D/H\@.  Thus the scales we
are considering enter the horizon after last scattering and they will induce
fluctuations in the CMB due to the inhomogeneities in the gravitational
potential from the variations in the baryon density, i.e.~the Sachs-Wolfe
effect, on scales $\theta\gtsim\theta_{\rm LS} \sim 2^\circ$.

The total temperature fluctuations in the CMB for isocurvature perturbations
including the intrinsic fluctuations and those induced by the Sachs-Wolfe
effect are
\be \frac{\delta T}{T} \approx 2\delta\phi, \ee
where $\delta\phi$ is the Newtonian potential (Hu \& Sugiyama~1995).
This expression is accurate to $\sim10\%$.
For isocurvature perturbations the fluctuations in the potential are 
\be \delta\phi  \sim \frac{G \delta M}\lambda \sim
 \frac12 H^2 \Omega_B \lambda^2 {\delta n_B \over n_B} \sim
 \frac23 \Omega_B \left( \frac{\rho_R}{\rho_B} \right)
\delta \label{df} \ee
at the time when a scale $\lambda \sim H^{-1}$ enters the horizon.  Here
\be \delta \equiv \frac{\delta\left( n_B/s \right)}{n_B/s} \sim
\frac34 {\rho_B \over \rho_R} {\delta n_B \over n_B} \sim
\frac{\delta\eta}\eta  \label{kt} \ee
is the fluctuation in the number of baryons per comoving volume and
$\rho_R$ is the energy density in radiation at late times. This expression
(\ref{kt}) is valid for $\rho_B \gg \rho_R$. The
temperature fluctuations are suppressed by two effects.  First by the density
of baryons in the Universe, $\Omega_B$, which create the potential well.  The
second suppression comes from the fact that super-horizon sized perturbations
cannot grow since $\delta\rho = 0$.

As one can see from Eq.(\ref{df}), the potential fluctuations $\delta \phi$ are 
time independent on a particular scale $\lambda$, prior to
horizon crossing as $\lambda^2 \delta n_B/n_B \propto R^3 \propto t^2$, and 
$H^2 \propto t^{-2}$.  However as can also be seen from (\ref{df}), 
these fluctuations  are scale dependent. Prior to horizon crossing, 
the suppression factor $(\rho_R/\rho_B)$ induces a scale dependence and
we can scale $\delta\phi$ to the present by noting that $\delta\phi \propto
R^{-1}$ and that the scale $\lambda$ enters the horizon at a time $t \propto
\lambda^3$.  Since $R\propto t^{2/3}$ in the matter dominated era
\be \delta\phi = \left( \frac{\lambda_0}\lambda \right)^2 \delta\phi_0 \ee
in terms of the potential fluctuations today, $\delta\phi_0$.  For $\lambda_0
\sim H_0^{-1}$ and $\delta \sim 1$ the restriction $\delta T/T \ltsim 10^{-5}$
leads to $\lambda \gtsim 2.4\lambda_0$.
This is our main result. The large scale, large amplitude
baryon inhomogeneity needed to explain the apparent dispersion in the D/H 
abundances measured in quasar absorption systems makes a 
contribution to the
microwave background anisotropy induced by the Sachs-Wolfe effect which is
in excess of the observed anisotropy for all scales $\lambda < \lambda_0 \sim
H_0^{-1}$ that have entered the horizon.

On angular scales somewhat smaller than 2$^\circ$, the Sachs-Wolfe effect is
not operative but a similar constraint may be derived.
For scales which enter the horizon between the epochs of matter domination and
last scattering, ie between $\sim 20 'h^{-1}$ and 2$^\circ$, 
corresponding to length scales between $\sim 35 h^{-2}$ Mpc and 200 $h^{-1}$ Mpc,
there is a contribution to $\delta T/T$ which is due to Doppler shifts across 
fluctuations at last scattering. This gives (Silk 1991)
\be {\delta T \over T} \sim {v \over c} \sim
H \Omega_B \lambda {\delta n_B \over n_B} \sim
  \Omega_B \delta \ee
Thus large fluctuation in $n_B/s$  are excluded at these scales as well.

To avoid these CMB constraints we need to consider smaller scales 
where the fuzziness of the last scattering surface suppresses temperature 
fluctuations so that at scales below $30 ''$ the fluctuations
are acceptably small.  Alternatively we can consider  scales that have not yet
been probed by the CMB\@.  However we do not want to consider scales that
are so small that many such perturbations would make up one quasar absorption
cloud.  In this case we would expect the regions to mix.  Observations in our
Galactic neighborhood preclude this mechanism from producing the large
fluctuations in the D/H abundance (Copi, Olive, \& Schramm~1995; Jedamzik \&
Fuller~1995).  Very large fluctuations mix in regions with high \li7\
abundances producing a final abundance inconsistent with present observations.
Fluctuations with $\delta \ltsim 0.15$ can be made consistent with the light
element abundances (Copi, Olive, \& Schramm~1995) but are not sufficiently
large to explain the observed QSO D/H abundance variations.

To circumvent both of these bounds one is forced to consider a scale in
 the middle; large enough to encompass an entire quasar absorption cloud, but
 small enough to avoid the CMB bounds. This leaves us only with a scale of
 $\sim 1\Mpc$ which corresponds to $\theta \sim 30''$ at last scattering.
 Extra-galactic \hii\ regions, the best sites for observing \he4\ have been
 probed on these scales.  Both \D\ and \li7\ have been probed in our Galactic
 neighborhood on scales smaller than this. Thus we would not expect to see
 variations in the observed Galactic \D\ and \li7\ abundances but would expect
to see  them in the observed \he4\ abundances.  Observations from the lowest
metallicity extra-galactic \hii\ regions are consistent with a scatter of
$\delta Y_p \sim 0.01$ which corresponds to $\delta\eta\sim3$.  However the
uncertainties are large and the scatter is consistent with the assigned error
bars of individual \he4 measurements.

We note that picking out a single scale for the baryon number fluctuations
strays far from our original motivation of flat directions from supersymmetry
which generally lead to a {\em flat} spectrum on very large scales.  Thus, we
cannot motivate the $30''$ scale from the model described above.  As
we noted above however, such a model may produce fluctuations with
significantly smaller amplitudes.  Baryon inhomogeneities of this type would
affect the ratio of the baryon mass to the total mass as measured by the x-ray
emission of hot gas in clusters of galaxies.

 In the model for baryon number fluctuations described above, though
the amplitude for isocurvature fluctuations  is contrained so that it may not
explain the possible dispersion in D/H, these fluctuations may be present with
a smaller amplitude. As such, these isocurvature
baryon number fluctuations are: 1) expected to have a flat spectrum; 2) are 
Gaussian distributed, ie they are random and; 3) most importantly, they are
completely
independent of the overall dark matter fluctuations (eg.\ due
to adiabatic fluctuations produced during inflation).
Thus, we would generally expect
a higher baryon to dark matter ratio in clusters than the Universal average.
Note that adiabatic fluctuations do not lead to these types of variations since
$\delta n_B/n_B = \delta n_{\rm CDM}/n_{\rm CDM}$ so the ratio remains
constant.  In fact, we would expect the ratio $M_{\rm hot gas}/M_{\rm tot}$ to
be different in different clusters which is consistent with the wide variations
see in current
observations (Mushotzky 1996, or 
see for example Wu \& Fang 1996), though one must be cautious
since different observations may not be directly comparable (Evrard, Metzler,
\& Navarro~1995).
 On the other hand, such variations are hard to understand in
traditional cold dark matter models, since cluster scales are supposed to be
fair samples of the overall matter distribution and it is assumed that there is
a common source for the fluctation spectrum. 

\bigskip

At present the observations of deuterium in quasar absorption systems roughly
fall around two values that differ by an order of magnitude.  Though it is
premature to identify either (or any) set of observations as primordial we
consider the possibility that both are correct and are primordial.  To be
primordial we need large scale, large amplitude isocurvature perturbations.
Standard BBN would then lead to the observed dispersion in D/H.  Supersymmetry
provides one model where the required isocurvature spectrum can be
produced.

	For scales that enter the horizon after last scattering, $\theta >
\theta_{\rm LS} \sim 2^\circ$, the dominant contribution to the CMB is the
Sachs-Wolfe effect.  Perturbations on all scales that have entered the horizon
after last scattering that could explain the D/H
dispersion lead to CMB fluctuations that are larger than observed.  For scales
smaller than this, $20'h^{-1} \ltsim \theta \sim 2^\circ$, Doppler shifts across
fluctuations at last scattering would also be too large for these
fluctuations.  Only on smaller scales can the CMB limits be avoided.  However
scales smaller than those of a quasar absorber would be well mixed thus
homogenizing the BBN products.  Fluctuations of the amplitude we are
considering here are not possible due to light element constraints (Copi,
Olive, \& Schramm~1995; Jedamzik \& Fuller~1995).  Only on scales $\lambda
\sim1\Mpc$, $\theta\sim30''$ can we avoid all constraints.  Observations of
\he4\ in extra-galactic \hii\ regions probe these scales but the data is too
uncertain to test for these perturbations.  Even so, we do not have a model
that will produce large amplitude perturbations on just these scales.

	Although large amplitude, large scale perturbations leave excessive
imprints on the CMB, smaller amplitude perturbations would not.  Smaller
amplitude perturbations cannot explain the D/H dispersion but can address the
cluster baryon problem.  Cluster scale isocurvature perturbations can explain
different $M_{\rm hotgas}/M_{\rm tot}$ ratios in clusters that cannot be
understood in the standard cold dark matter scenarios.

\section*{Acknowledgments}

	We would like to thank M. Srednicki, E. Vishniac, and M. White for
useful conversations.  
This work has been supported in part by NSF grant AST 90-22629, DOE grant
DE-FG02-91-ER40606, and NASA grant NAGW-1321 at the
University of Chicago, by the DOE and by NASA through grant NAGW-2381 at
Fermilab, and the U. S. Department of Energy under contract
DE-FG02-94ER-40823 at the University of Minnesota.

\endpaper

\beginapjbib

\bibitem Affleck, I. \& Dine, M. 1985, Nucl. Phys.,  B249, 361.

\bibitem Balser, D.S., Bania, T.M., Brockway, C.J.,
Rood, R.T., \& Wilson, T.L. 1994, ApJ, 430, 667.


\bibitem Boothroyd, A.I. \& Malaney, R.A. 1995, astro-ph/9512133.

\bibitem Boothroyd, A.I. \& Sackman, I.-J. 1995, astro-ph/9512121.

\bibitem Burles, S. \& Tytler, D. 1996, astro-ph/9603070.

\bibitem Cardall, C.Y. \& Fuller, G.M. 1996, astro-ph/9603071.

\bibitem Carswell, R.F., Rauch, M., Weymann, R.J., Cooke, A.J. \&
Webb, J.K. 1994, MNRAS, 268, L1.

\bibitem  Carswell, R.F., \etal. 1996, MNRAS, 278, 518.


\bibitem Charbonnel, C. 1994, A \& A, 282, 811.

\bibitem Charbonnel, C. 1995, ApJ, 453, L41.

\bibitem Copi, C.J., Olive, K.A., \& Schramm, D.N. 1995, ApJ, 451, 51.

\bibitem Copi, C.J., Schramm, D.N., \& Turner, M.S.~1995,  Science, 267,
192.

\bibitem Dearborn, D., Steigman, G. \& Tosi, M. 1996, ApJ, 465, in press.

\bibitem Ellis, J., Enqvist, K., Nanopoulos, D.V. \& Olive, K.A. 1987, 
Phys. Lett.,  B191, 343.

\bibitem Epstein, R., Lattimer, J., \& Schramm, D.N. 1976, Nature, 263, 198.


\bibitem Evrard, A.E., Metzler, C.A., \& Navarro, J.F.~1995, astro-ph/9510058.

\bibitem Fields, B.D. \& Olive, K.A. 1996, Phys Lett B368, 103.

\bibitem Fields, B.D., Kainulainen, K., Olive, K.A., \& Thomas, D. 1996
New Astronomy, in press.

\bibitem Fuller, G.M. \& Cardall, C.Y. 1996, astro-ph/96006025.

\bibitem Galli, D., Palla, F. Ferrini, F., \& Penco,U. 1995, ApJ,
433, 536.

\bibitem Geiss, J. 1993, in {\it Origin
 and Evolution of the Elements} eds. N. Prantzos, 
E. Vangioni-Flam, \& M. Cass\'{e}
(Cambridge: Cambridge University Press), p. 89.

\bibitem Gloeckler, G. \& Geiss, J. 1996, Nature, 381, 210.


\bibitem Hata, N., Stiegman, G., Bludman, S., \& Langacker, P. 1996,
astro-ph/9603087.

\bibitem Hu, W. \& Sugiyama, N.~1995 PRD, 51, 2599.


\bibitem Hogan, C.J. 1995, ApJ, 441, L17.

\bibitem Jedamzik,~K. \& Fuller,~G.M.~1995, ApJ, 452, 33.

\bibitem Kurki-Suonio,~H., Jedamzik,~K., \& Mathews,~G.J.~1996,
astro-ph/9606011.

\bibitem Linde, A.D. 1985a, Phys. Lett., B160, 243.

\bibitem Linde, A.D. 1985b, Phys. Lett., B158, 375.

\bibitem Malaney,~R.~A. \& Mathews,~G.~1993,  Phys. Rep., 229, 147.

\bibitem Mushotzky, R.F. 1996, talk given at the Berkeley Conference on 
the Baryonic Content of the Universe.

\bibitem Olive, K.A., Rood, R.T., Schramm, D.N., Truran, J.W.,
\& Vangioni-Flam, E. 1995, ApJ, 444, 680.

\bibitem Olive, K.A., \& Scully, S.T. 1996, Int. J. Mod. Phys., A11,
409.

\bibitem Olive, K.A., \& Steigman, G. 1995, ApJS, 97, 49.





\bibitem Reeves, H., Audouze, J., Fowler, W.A., \& Schramm, D.N. 1973, 
ApJ, 179, 909.

\bibitem Rood, R.T., Bania, T.M., \& Wilson, T.L. 1992, Nature, 355, 618.

\bibitem Rood, R.T., Bania, T.M.,  Wilson, T.L., \& Balser, D.S. 1995, 
in {\it
 the Light Element Abundances, Proceedings of the ESO/EIPC Workshop},
ed. P. Crane, (Berlin:Springer), p. 201.

\bibitem Rugers, M. \& Hogan, C. 1996a, ApJ,  259, L1.

\bibitem Rugers, M. \& Hogan, C. 1996b, in {\it Cosmic Abundances},
proceedings of the 6th Annual October Astrophysics Conference in Maryland,
PASP conference series, in press.

\bibitem Rugers, M. \& Hogan, C. 1996c, astro-ph/9603084.

\bibitem Scully, S.T., Cass\'{e}, M., Olive, K.A., Schramm, D.N., 
Truran, J., \& Vangioni-Flam, E. 1996, ApJ, 462, 960.

\bibitem Silk, J. 1991, Phys. Scr., 36, 16.




\bibitem Schramm, D.N. \& Turner, M.S. 1996, Nature, 381,193.

\bibitem Songaila, A., Cowie, L.L., Hogan, C. \& Rugers, M. 1994
Nature, 368, 599.

\bibitem  Steigman, G. 1994, MNRAS, 269, L53.

\bibitem Steigman, G., Fields, B. D., Olive, K. A., Schramm, D. N.,
\& Walker, T. P., 1993, ApJ 415, L35.

\bibitem Thomas,~D., Schramm,~D.~N., Olive,~K.~A., Mathews,~G.~J., Meyer,~B.~S.
\& Fields,~B.~D.~1994,  ApJ, 406, 569.


\bibitem Turner, M.S., Truran, J.W., Schramm, D.N., \& Copi, C.J. 1996,
astro-ph/9602050.

\bibitem Tytler, D. \& Burles, S. 1996, astro-ph/9606110.

\bibitem  Tytler, D. \& Fan, X.-M. 1995,  BAAS, 26, 1424.

\bibitem Tytler, D., Fan, X.-M., \& Burles, S. 1996, astro-ph/9603069.

\bibitem Vangioni-Flam, E. \& Cass\'{e}, M. 1995, ApJ, 441, 471.

\bibitem Walker,~T.~P., Steigman,~G., Schramm,~D.~N., Olive,~K.~A., \&
Kang,~H.~1991,  ApJ, 376, 51.

\bibitem Wampler, E.J. \etal 1996, astro-ph/9512084, AA, in press.

\bibitem Wampler, E.J. 1996, scientific correspondence submitted to Nature.

\bibitem Wasserburg, G.J., Boothroyd, A.I., \& Sackmann, I.-J. 1995, ApJ,
447, L37.

\bibitem Weiss, A., Wagenhuber, J., \& Denissenkov, P. 1995, astro-ph/9512120.

\bibitem Wu, X.-P. \& Fang, L.-Z. 1996, astro-ph/9606109.


\bibitem Yokoyama, J. 1994, AstroParticle Phys, 2, 291.
\endapjbib

\newdimen\boxdim
\setbox0=\hbox{$\Box$}
\boxdim=\ht0

\begin{figure} \center\leavevmode \epsfysize=\hsize
\rotate[r]{\epsfbox{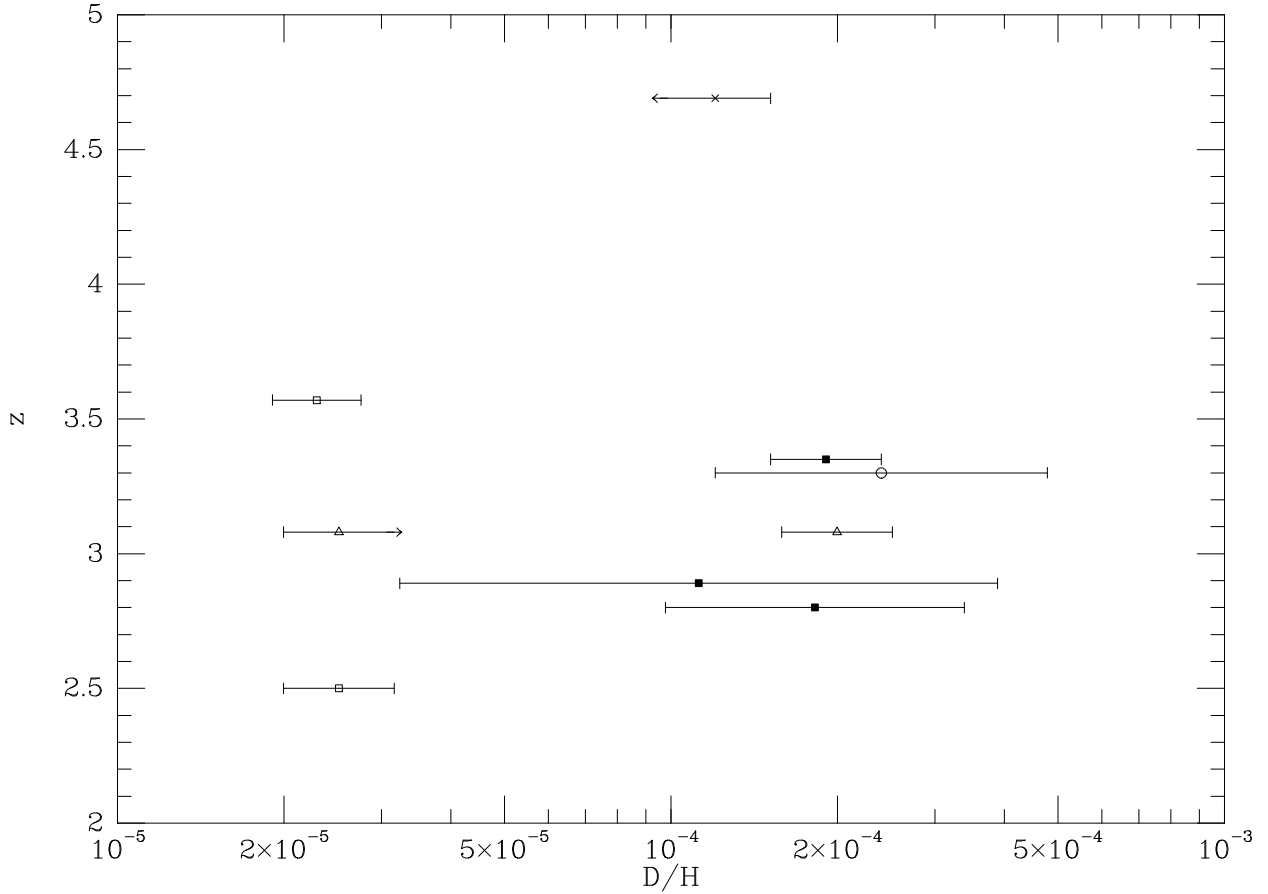}}
\caption{A summary of the D/H measurements in quasar absorption
systems.  The observations were made by ($\circ$)~Carswell \etal~(1994) and
Songaila \etal~(1994), (\vrule height\boxdim width\boxdim depth0pt)~Rugers \&
Hogan~(1996a,b,c), ($\bigtriangleup$)~Carswell \etal~(1996), ($\times$)~Wampler
\etal~(1996), and ($\Box$)~Tytler, Fan, \& Burles~(1996) and Burles \&
Tytler~(1996).  The Rugers \& Hogan (1996a) 
 (\vrule height\boxdim width\boxdim depth0pt) point and the ($\circ$) point
have been separated in redshift for clarity.
  See the text for general details.}
\end{figure}

\end{document}